 \documentclass[preprint,12pt]{elsarticle}
 \pdfoutput=1

\usepackage{amsmath}
\usepackage{color}
\usepackage{amssymb}
\usepackage{psfrag}
\usepackage{listings}
\usepackage[hyphens]{url}
\usepackage[hidelinks]{hyperref}
\hypersetup{breaklinks=true}
\usepackage{cite}

\newcommand{\diff}{\mathrm{d}}
 \newcommand{\iuw}{I}
 \newcommand{\juw}{\nu}
\DeclareMathOperator{\sign}{sign}

\journal{Computers \& Fluids}

\begin{document}

\begin{frontmatter}

\title{On algebraic TVD-VOF methods for \\ tracking material interfaces}

%% use optional labels to link authors explicitly to addresses:
%% \author[label1,label2]{<author name>}
%% \address[label1]{<address>}
%% \address[label2]{<address>}

\author[1]{Sergio Pirozzoli}
\ead{sergio.pirozzoli@uniroma1.it}
\author[1]{Simone Di Giorgio}
\author[2]{Alessandro Iafrati}
%\cortext[aut1]{Corresponding author: sergio.pirozzoli@uniroma1.it}
%\ead{+390644585202}
\address[1]{Dipartimento di Ingegneria Meccanica e Aerospaziale, Sapienza Universit\`a di Roma, via Eudossiana 18, 00184 Rome, Italy}
\address[2]{CNR-INSEAN, Istituto Nazionale per Studi ed Esperienze di Architettura Navale, Via di Vallerano 139, 00128 Rome, Italy}

\begin{abstract}
We revisit simple algebraic VOF methods for advection of material interfaces based of the
well established TVD paradigm. We show that greatly improved representation of 
contact discontinuities is obtained through use of a novel CFL-dependent limiter whereby
the classical TVD bounds are exceeded. Perfectly crisp numerical interfaces are obtained
with very limited numerical atomization (flotsam and jetsam) as compared to previous
SLIC schemes. Comparison of the algorithm with accurate geometrical VOF shows 
larger error at given mesh resolution, but comparable efficiency 
when the reduced computational cost is accounted for.
\end{abstract}

\begin{keyword}
%% keywords here, in the form: keyword \sep keyword

%% MSC codes here, in the form: \MSC code \sep code
Material interfaces \sep Multiphase flows \sep VOF method \sep TVD schemes
%% or \MSC[2008] code \sep code (2000 is the default)

\end{keyword}

\end{frontmatter}

%%
%% Start line numbering here if you want
%%
% \linenumbers

%% main text
\section{Introduction}
\label{sec:introduction}

Passive advection of sharp interfaces is an important and difficult problem
in computational physics. Let $\chi$ be a passive tracer advected by a continuous
divergence-free velocity field ${\bf u}$, it satisfies the transport equation
\begin{equation}
\frac{\partial \chi}{\partial t} + \nabla \cdot (\chi {\bf u}) = 0. \label{eq:scalar}
\end{equation}
The problem under scrutiny here consists of the case that $\chi$ is either $1$ or $0$,
corresponding to the case of two immiscible fluids.
Although seemingly harmless, the problem is notoriously difficult for numerical methods,
mainly because it requires algorithms which must be capable of capturing 
discontinuities without allowing spreading of the interface nor violation of the
range-preservation conditions. 
These restrictions make it difficult to apply classical shock-capturing methods 
of computational gas dynamics~\citep{pirozzoli_11}, which work well at shocks
on account of nonlinear self-focusing, but which are not optimal for contact discontinuities
which are unavoidably spread in time. Altough corrections to classical shock-capturing 
schemes have been proposed to improve the behavior at contacts~\citep{harten_77,yang_90},
these are typically computationally intensive, and include adjustable parameters.
 
The literature on numerical algorithms for advection of material interface is nowadays immense,
and partly summarized in reference textbooks~\citep{tryggvason_11}. Although many alternatives are available,
it appears the the volume-of-fluid (VOF) approach, in which the cell averages of $\chi$
are evolved in time to approximate Eqn.~\eqref{eq:scalar}, 
is the most popular mainly because of its simplicity 
and built-in discrete mass conservation.
An exact VOF transport algorithm does in fact exist for binary functions~\citep{tryggvason_11} 
in one space dimension. However, when the exact algorithm is extended to 
multiple space dimensions in direction-wise fashion~\citep{noh_76,hirt_81},
in addition to distorting the interfaces, it also generates considerable 
amount of 'floatsam and jetsam', consisting of pieces of the interface breaking away in unphysical way.
Shortcomings of the exact transport algorithm have been traditionally cured 
by introducing multi-dimensional information, expecially through geometrical reconstruction 
of the interface within interface computational cells, followed by exact advection. 
This is the essence of the PLIC (piecewise linear interface calculation) 
approach~\citep{youngs_82,scardovelli_03,pilliod_04} which in its disparate variants
is currently regarded to be the best option for VOF algorithms, and it is implemented in 
popular public domain solvers~\citep{GERRIS, Basilisk}.
The main drawback of geometric VOF methods with respect to those which do not require geometrical reconstruction
(hereafter referred to as `algebraic' methods) is clearly higher coding complexity
and computational cost, to an extent which we will attempt to quantify later on.
Hence, algebraic algorithms are still occasionally used 
as a cheap alternative to PLIC. Amongst the various attempts it is worth quoting the 
THINC (tangent hyperbola for interface capturing) method~\citep{xiao_05}, based on use of
hyperbolic tangent basis functions to approximate solutions with jumps.
It is the goal of this paper to further explore the predictive capabilities of algebraic methods
by revising TVD shock-capturing schemes~\citep{vanleer_74,sweby_84},
based on locally linear recostruction of the numerical solution in each cell
with suitably limited slope.
We will show that judicious choice of the slope limiters yields a new class of schemes
which strictly maintain sharp interfaces in time, preserve the range-preservation condition,
and limit interface wrinkling with subsequent formation of floatsam and jetsam.
The algorithm is extremely simple to implement as it wholly avoids multi-dimensional reconstructions,
which makes it competitive in terms of efficiency with higher-order geometric VOF methods.

\section{TVD reconstruction}
\label{sec:numerics}

In the VOF method a color function is introduced to approximate the cell average of $\chi$, 
which in the illustrative case of one space dimension is defined as
\begin{equation}
C_i^n = \frac 1{\Delta x_i} \int_{x_{i-1/2}}^{x_{i+1/2}} \chi (x,t^n) \diff x, \label{eq:color}
\end{equation}
where $\Delta x_i = x_{i+1/2}-x_{i-1/2}$ is the cell length, and $n$ is the time index.
Equation~\eqref{eq:scalar} is then discretized in its integral form,
leading to
\begin{equation}
C_i^{n+1} = C_i^n - \frac 1{\Delta x_i} \left( \hat{f}_{i+1/2} -\hat{f}_{i-1/2} \right), \label{eq:VOF}
\end{equation}
where the numerical flux $\hat{f}_{i+1/2}$ is an approximation for the amount of $\chi$ which is 
transported through the cell interface $x_{i+1/2}$ during the time interval $(t^n, t^{n+1})$.
In incompressible Navier-Stokes solvers the velocity components typically come in 
a staggered arrangement~\citep{harlow_65,orlandi_12}, hence $u_{i+1/2}$ is known point-wise
at the cell faces.
According to classical upwinding arguments, a piece-wise linear reconstruction of the color function is considered
either in the left neighboring cell ($\iuw=i$) if $u_{i+1/2} \ge 0$,
or in the right cell ($\iuw=i+1$) if $u_{i+1/2} < 0$,
\begin{equation}
C_{\iuw}(x) = C^n_{\iuw} + s_{\iuw} (x-x_{\iuw}). \label{eq:linear}
\end{equation}
The slope of the reconstructed color function is then selected so as to 
prevent the occurrence of overshoots/undershoots by enforcing the TVD constraints~\citep{sweby_84},
\begin{equation}
s_{\iuw} = \frac 1{\Delta x_{\iuw}} \, \varphi(\theta_{i+1/2}) \, \delta C_{i+1/2} , \label{eq:slope}
\end{equation}
where $\delta C_{i+1/2} = C_{i+1}-C_i$, $\theta_{i+1/2}=\delta C_{i-\juw+1/2}/\delta C_{i+1/2}$,
with $\juw=\sign u_{i+1/2}$, and $\varphi$ a suitable slope limiter function.
Non-oscillatory TVD reconstructions are obtained provided the following restrictions 
for $\varphi$ are satisfied
\begin{equation}
0 \le \varphi \le \frac 2{1-\sigma_{i+1/2}}, \quad 0 \le \varphi \le \frac {2 \theta_{i+1/2}}{\sigma_{i+1/2}}, \quad \forall i, \label{eq:TVD} 
\end{equation}
where $\sigma_{i+1/2} = \vert u_{i+1/2} \vert \Delta t / \Delta x_{\iuw}$ is the local Courant number
($0 \le \sigma_{i+1/2} \le 1$ for time stability), with $\Delta t = t^{n+1}-t^n$.
The TVD constraints \eqref{eq:TVD} are typically enforced in the stronger sufficient form
\begin{equation}
0 \le \varphi \le 2, \quad 0 \le \varphi \le 2 \theta_{i+1/2}, \quad \forall i, \label{eq:TVD2}
\end{equation}
which applies irrespective of the local Courant number.
Based on the reconstruction~\eqref{eq:linear}, straightforward time integration
yields the numerical flux
\begin{equation}
\hat{f}_{i+1/2} = u_{i+1/2} \left( C^n_{\iuw} + \frac {\juw}2 \left( 1 - \sigma_{i+1/2} \right) \varphi (\theta_{i+1/2}) \delta C_{i+1/2} \right) . \label{eq:flux}
\end{equation}
A large number of slope limiters have been proposed in the literature, yielding
numerical schemes with vastly different properties, some of which were reviewed by \citet{kemm_11}.
Most classical limiters have $\varphi(1)=1$, which guarantees second-order accuracy for smooth solutions.
However, this choice is unnecessary in the case of sharp material interfaces 
in which the solution is non-smooth, and yields numerical diffusion of the interface in time.
More compressive schemes, which are rather prone to numerical squaring of smooth profiles 
are obtained by moving toward the upper boundary of the TVD
region defined by \eqref{eq:TVD}. Specifically, the least restrictive limiter is
\begin{equation}
\varphi_{UB} = \max \left(0, \min \left( \frac 2{1-\sigma_{i+1/2}}, \frac {2 \theta_{i+1/2}}{\sigma_{i+1/2}} \right) \right), \label{eq:UB}
\end{equation}
which we will refer to as upper-bound (UB) limiter, and sometimes
referred to as ultra-bee limiter~\citep{toro_13}. 
It is noteworthy that the upper horizontal bound of the limiter returns the first-order
downwind scheme, which is highly numerically unstable in a linear setting.
Hence, it is clear that the effect of limiting is to impart nonlinear stability 
to a scheme which would otherwise be unstable, and the favourable properties of limiter~\eqref{eq:UB}
for interface tracking results in competition between instability-driven 
numerical amplification and nonlinear bounds.
As shown by several authors~\citep{davis_94,despres_01}, the UB limiter yields a 
scheme which is identical to the exact interface transport solver in one space dimension.
Hence, as the latter, it yields excessive interface wrinkling in multidimensional problems.
A semi-discrete version of the UB limiter was considered by \citet{sweby_84},
\begin{equation}
\varphi_{SW} = \max \left(0, \min \left( 2 \theta_{i+1/2}, 2 \right) \right), \label{eq:SW}
\end{equation}
which also has strongly compressive behavior.
\begin{figure}
 \centering
 \includegraphics[]{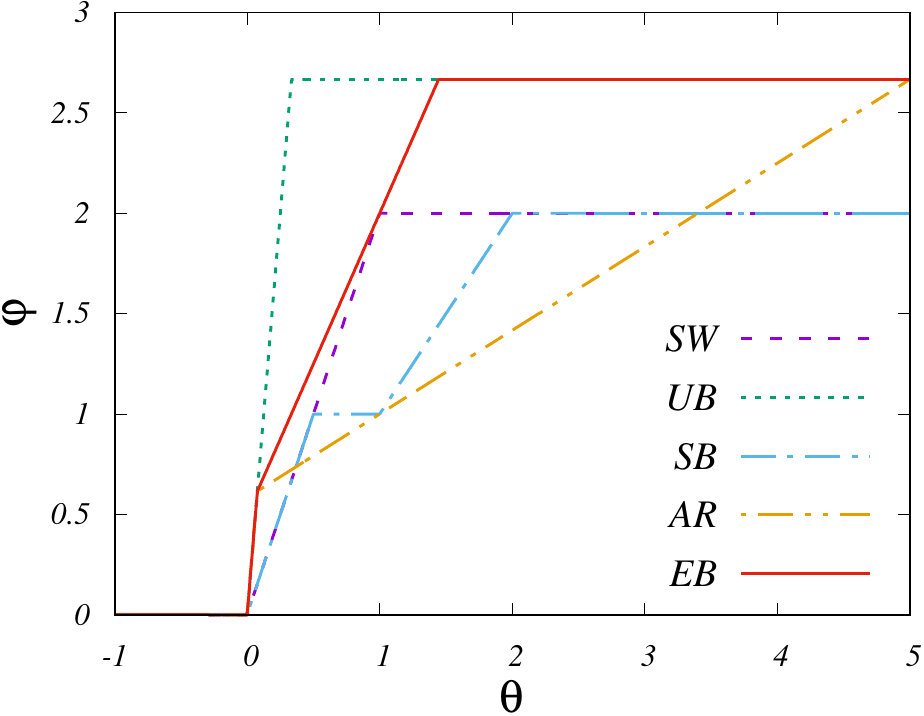} 
 \caption{Sweby diagram for the TVD limiters under scrutiny ($\sigma=0.25$ is assumed, and $s=3/2$).}
 \label{fig:limiters}
\end{figure}
A correction to the UB limiter was considered by \citet{arora_97}, designed in such a way as
to recover third-order accuracy in smooth regions,
\begin{equation}
\varphi_{AR} = \max \left(0, \min \left( \frac 2{1-\sigma_{i+1/2}}, 1+ \frac{1+\sigma_{i+1/2}}3 \left( \theta_{i+1/2} - 1 \right), \frac {2 \theta_{i+1/2}}{\sigma_{i+1/2}} \right) \right), \label{eq:AR}
\end{equation}
Another popular, rather compressive second-order limiter is Roe's super-bee~\citep{roe_84}, 
here considered in its semi-discrete version
\begin{equation}
\varphi_{SB} = \max \left(0, \min \left( 2 \theta_{i+1/2}, 1 \right), \min \left( \theta_{i+1/2}, 2 \right) \right). \label{eq:SB}
\end{equation}
As shown in the later discussion, we have found that effective limiters for 
transport of material interfaces must lie in the intermediate range between the SW and the UB limiters.
Hence, we propose the following new class of 'extra-bee' limiters
\begin{equation}
\varphi_{EB} = \max \left(0, \min \left( \frac 2{1-\sigma_{i+1/2}}, \frac {2 \theta_{i+1/2}}{\sigma_{i+1/2}}, 2 + s (\theta_{i+1/2}-1) \right) \right). \label{eq:EB}
\end{equation}
Based on a series of numerical tests, we have empirically found that a nearly optimal
value for the adjustable slope which appears in \eqref{eq:EB} is $s=3/2$, which is thus 
retained in all later results.
The Sweby $(\theta,\varphi)$ diagrams for the TVD limiters considered so far are 
plotted in Fig.~\ref{fig:limiters}, in which the time-dependent limiters are shown 
for $\sigma_{i+1/2}=0.25$.

%Improved accuracy with respect to non-uniform transport velocity can be gained assuming
%linear variation of $u$ within each cell, with slope $t_i = (u_{i+1/2}-u_{i-1/2})/\Delta x_i$. The higher-order (in $u$) counterpart of the first-order flux \eqref{eq:flux} is
%\begin{equation}
%\hat{f}_{i+1/2} = u_{i+1/2} \left\{ C_{\iuw} \left( 1 - \frac 12 t_{\iuw} \Delta t \right) + \frac {\juw}2 \left[ \left( 1 - \sigma_{i+1/2} \right) - \frac 16 t_{\iuw} \Delta t \left(3-4 \sigma_{i+1/2} \right) \right] \varphi (\theta_{i+1/2}) \delta C_{i+1/2} \right\}, \label{eq:fluxHO}
%\end{equation}
%which comes at a marginally higher computational cost.
The extension to multiple space dimensions is obtained through straightforward 
direction-wise application of the numerical flux~\eqref{eq:flux}. % (or \eqref{eq:fluxHO}).
We have found that much greater robustness is obtained through use of 
Strang's time splitting procedure~\citep{strang_68}, whereby the numerical solution is updated
after each direction-wise sweep. In order to preserve good numerical isotropy it is also important 
that the order of the space sweeps is exchanged between consecutive time steps. 
Although multi-dimensional limiting is possible and has been exploited with some success~\citep{zalesak_79}
here we refrain from any attempt to construct a genuinely multi-dimensional scheme, and 
explore the predictive capabilities of TVD algorithms within the standard 
direction-wise setting.

\section{Results}
\label{sec:results}

Numerical results of representative TVD schemes are hereafter reported for one- and multi-dimensional
benchmark problems. For reference purposes, the results obtained with the THINC algebraic transport 
scheme and with the public domain Basilisk solver (as a computationally efficient representative of the PLIC method) are also shown. 
Regarding the THINC algorithm we have found that the baseline implementation given in
\citet{xiao_05} yields poor results for certain test cases, hence
we present results of the THINC/SW variant, which incorporates a simple multi-dimensional 
correction accounting for the local interface orientation~\citep{xiao_11}.
For the sake of uniformity, all the numerical simulations have been
carried out at global Courant number (i.e. based on the maximum advection speed) of $0.25$.

\subsection{One-dimensional advection}

\begin{figure}
 \centering
 \includegraphics[]{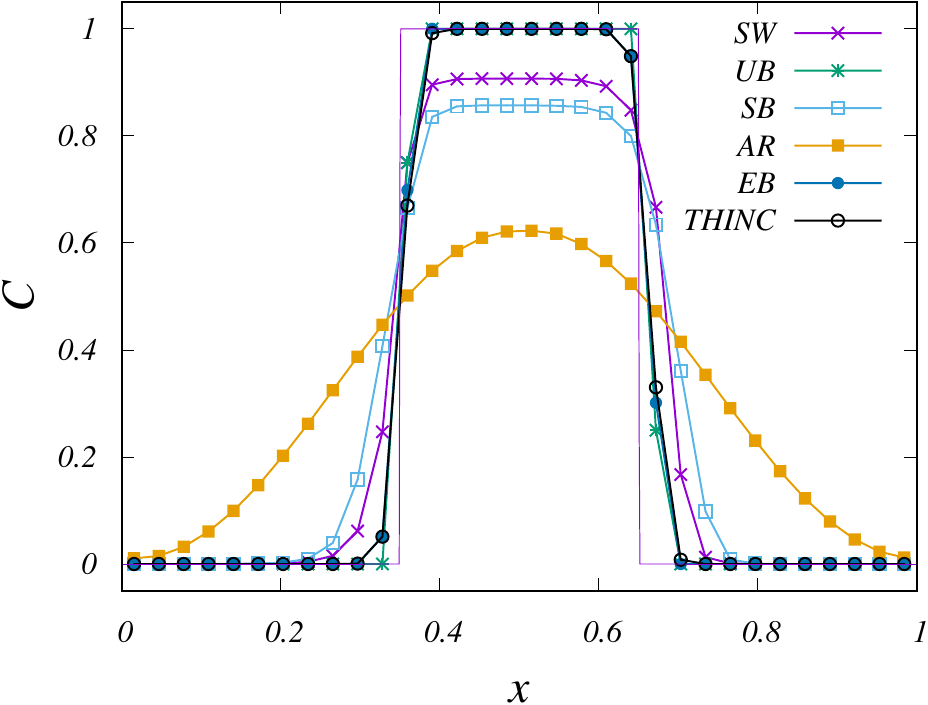} 
 \caption{One-dimensional advection of top-hat wave. Results are presented with $N_x=32$ grid points, after 100 advection periods. The thin solid line denotes the reference solution.} 
 \label{fig:1d}
\end{figure}

As a preliminary test, the TVD-VOF solver was applied to the one-dimensional transport of a top-hat wave
with unit advection velocity, in a $[0,1]$ periodic domain discretized with $32$ cells. 
The numerical solution was evolved in time for 100 periods, allowing to clearly discern 
possible issues. The computed solutions are shown in Fig.~\ref{fig:1d}, where we also report
results obtained with the THINC algorithm. 
We must note that common compressive limiters as SW and SB do in fact diffuse the wave in the long run,
whereas the solution range is only exactly preserved by the UB and EB limiters, which 
are capable of retaining the same discrete profile for infinitely long time.
Slight differences between the two limiters may however be noticed,
as the UB limiter resolves the interface with only one grid point in the transition zone, 
whereas the EB limiter requires two points.
Interestingly, the EB limiter yields a profile which is nearly identical 
to that given by THINC, in this one-dimensional test case.
The existence of asymptotic traveling wave solutions for TVD schemes
was discussed by \citet{roe_85}, who showed that monotonic 
exponential profiles can be retained for infinitely long time under certain conditions, 
and in particular the SB limiter supports a discrete profile including about 8-9 transition points.
%namely the same discrete profile is retained for infinitely long time~\citep{jiang_98}. 
The reason that the same behavior is not recovered in the present test is that
the square wave only contains about ten points, hence the head and the tail of the
wave interfere with each other. We have in fact verified that increasing the number 
of points to 64, the SB limiter also exhibits a traveling wave profile. 

\subsection{Zalesak slotted disk}

\begin{figure}[]
 \begin{center}
  \includegraphics[]{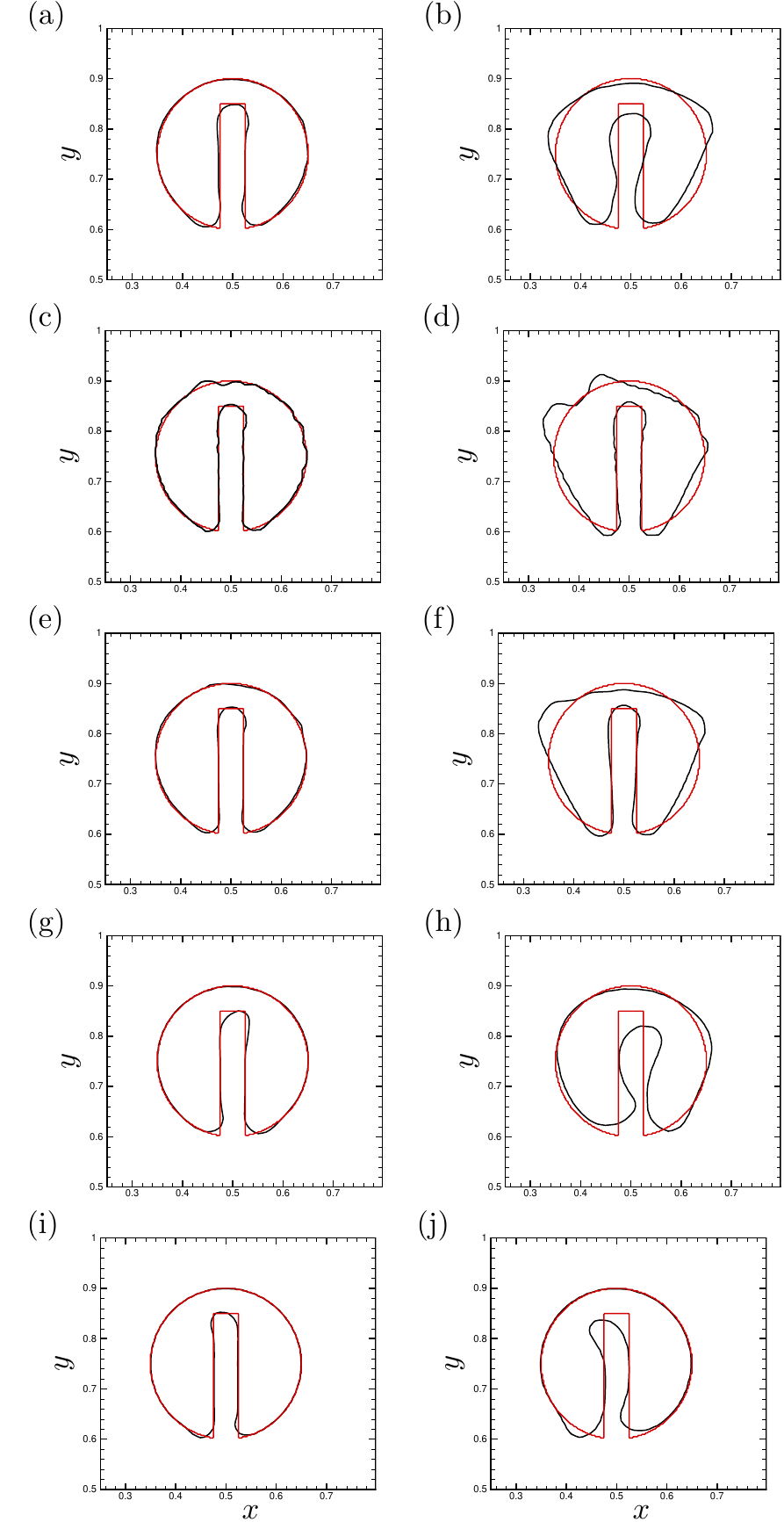} 
  \caption{Zalesak test case: $C=0.5$ iso-line after one revolution (left column) and after ten revolutions (right).
    Results are shown for methods SW-TVD (a-b), UB-TVD (c-d), EB-TVD (e-f), THINC/SW (g-h), PLIC (i-j).
    Black lines denote the numerical solution, and red lines the exact solution.}
  \label{fig:zalesak}
 \end{center}
\end{figure}

\begin{figure}
 \centering
 \includegraphics[]{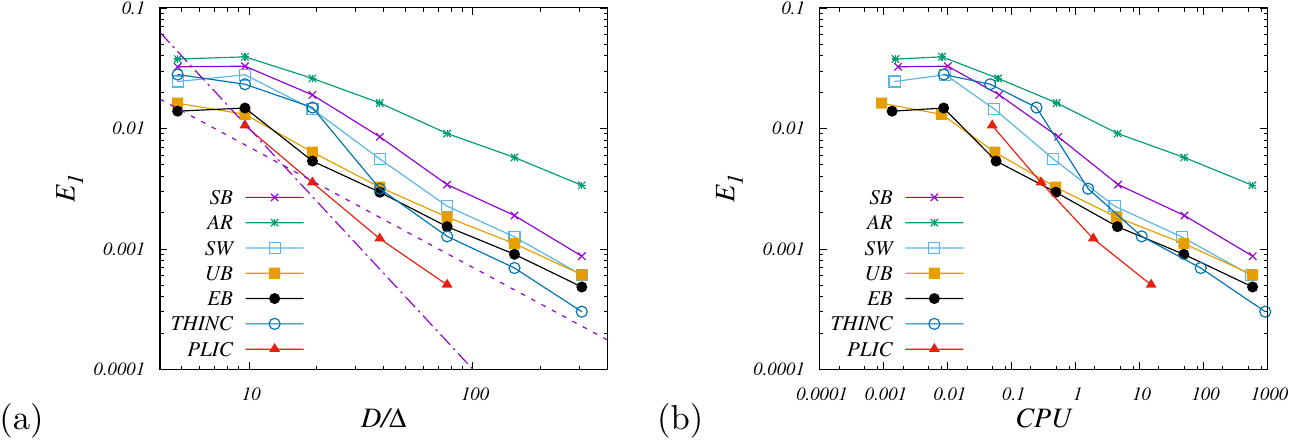} 
 \caption{Zalesak test case: $L_1$ error norm at $t=1$ as a function of mesh resolution (a), and as a function of CPU time (in seconds). The dashed and dash-dotted lines in panel (a) denote the $(D/\Delta)^{-1}$ and $(D/\Delta)^{-2}$ trends, respectively, with $D$ the disk diameter and $\Delta$ the mesh spacing.}
 \label{fig:err_Zalesak}
\end{figure}

This is a two-dimensional test case in which a circle with diameter $D=0.3$ and a $0.05$ wide slot is centered is centered at $(0.50, 0.75)$ in a unit square computational domain~\citep{zalesak_79}. The tracer is passively advected by a solid-body rotation velocity field defined by 
\begin{equation}
\left\{
\begin{array}{ccc} 
u & = & - 2 \pi \left( y - 0.5 \right) , \cr
v & = & \,\,\,\,  2 \pi \left( x - 0.5 \right) .
\end{array}
\right.
\end{equation}
Under these conditions the slotted disk rotates about the center of the 
domain, completing one full revolution every time unit.
Numerical results are shown in Fig.~\ref{fig:zalesak} for a 
$128^2$ uniform mesh, at $t=1$ and $t=10$.
For clarity of illustration, hereafter the $C=0.5$ iso-line is used to provide
a visual impression for the computed shape of the interface, which is by the way always 
resolved with 2-3 grid points at most.
At early times (left column) TVD-VOF schemes yield rather accurate representation of the
disk, although wiggles are observed around its periphery. Good representation of the inner slot
is recovered, especially with the EB limiter, which is not surprising given the
purely Cartesian nature of the solver. The THINC/SW solver yields smoother representation of the
outer perimeter, but (at least at this mesh resolution), it returns poorer approximation of the slot,
with more significant rounding of sharp edges. Similar conclusions apply to PLIC, 
although the error is even smaller in that case.
At later times (right column) TVD-VOF schemes still retain improved representation of the inner slot
(at least when the most aggressive limiters are applied), whereas the outer perimeter is severely
distorted, which again is expected given the strongly anisotropic nature of the solver.
The THINC/SW method has a more isotropic behavior, but it severely distorts the slot.
PLIC here shows clear superiority, for given mesh resolution. 

A quantitative error analysis for this test case is reported in Fig.~\ref{fig:err_Zalesak}.
In the left-hand-side frame the $L_1$ error norm, defined as
\begin{equation}
E_1 = \frac 1{N_x N_y} \sum_{ij} \vert C_{ij} - \tilde{C}_{ij} \vert, \label{eq:erl1}
\end{equation}
where $N_x$, $N_y$ are the number of cells in the coordinate directions, and $\tilde{C}$ is the exact solution,
is shown as a function of the mesh resolution expressed in terms of number of cells per disk diameter, $D/\Delta$. 
Meshes with $16^2$ to $1024^2$ cells are considered.
Given the presence of sharp corners, all methods under scrutiny exhibit first-order convergence.
Among the TVD-VOF methods, the EB limiter achieves the least error over the entire resolution range.
THINC/SW has higher error than EB-TVD at coarse resolutions, but slightly lower error at fine resolutions.
PLIC has lower error than other best schemes by a factor of about three, for given mesh resolution.
The higher accuracy of geometric VOF comes of course at higher computational cost than algebraic schemes, 
and this should be taken into account in the analysis. 
In order to carry out a quantitative computational efficiency analysis,
in Fig.~\ref{fig:err_Zalesak}(b) we show the $L_1$ error norm 
as a function of the computational cost of each simulation, 
which we quantify as the CPU time measured on a single Intel Core i7-6950X with 3.00GHz clock frequency.
Similar relative cost figures were by the way obtained using different core architectures. 
In general terms, we find that the typical grind time (i.e. CPU time per grid point per time step) 
of all presented TVD-VOF schemes is similar, whereas in these two-dimensional experiments 
THINC/SW is about a factor two more expensive
(mainly because of the large number of transcendental floating point operations), and
PLIC is at least a factor four more expensive.
In our interpretation, a scheme is more efficient is it can yield
the same accuracy for lower computational effort, or if it can provide
more accurate solutions for given cost.
The computational efficiency maps of Fig.~\ref{fig:err_Zalesak}(b) then lead the main conclusion that  
Cartesian VOF methods may be competitive with PLIC if relatively coarse representations are sufficient,
with relative error higher than about $0.3\%$, and requiring modest computational effort.
PLIC however retains clear superiority when stricter error tolerance is placed and/or 
more computational resources are available.
Among algebraic methods, we find that THINC/SW is only competitive with EB-TVD in the 
asymptotic high-resolution range.

\subsection{Rider-Kothe reversed single vortex flow}

\begin{figure}[]
 \begin{center}
  \includegraphics[]{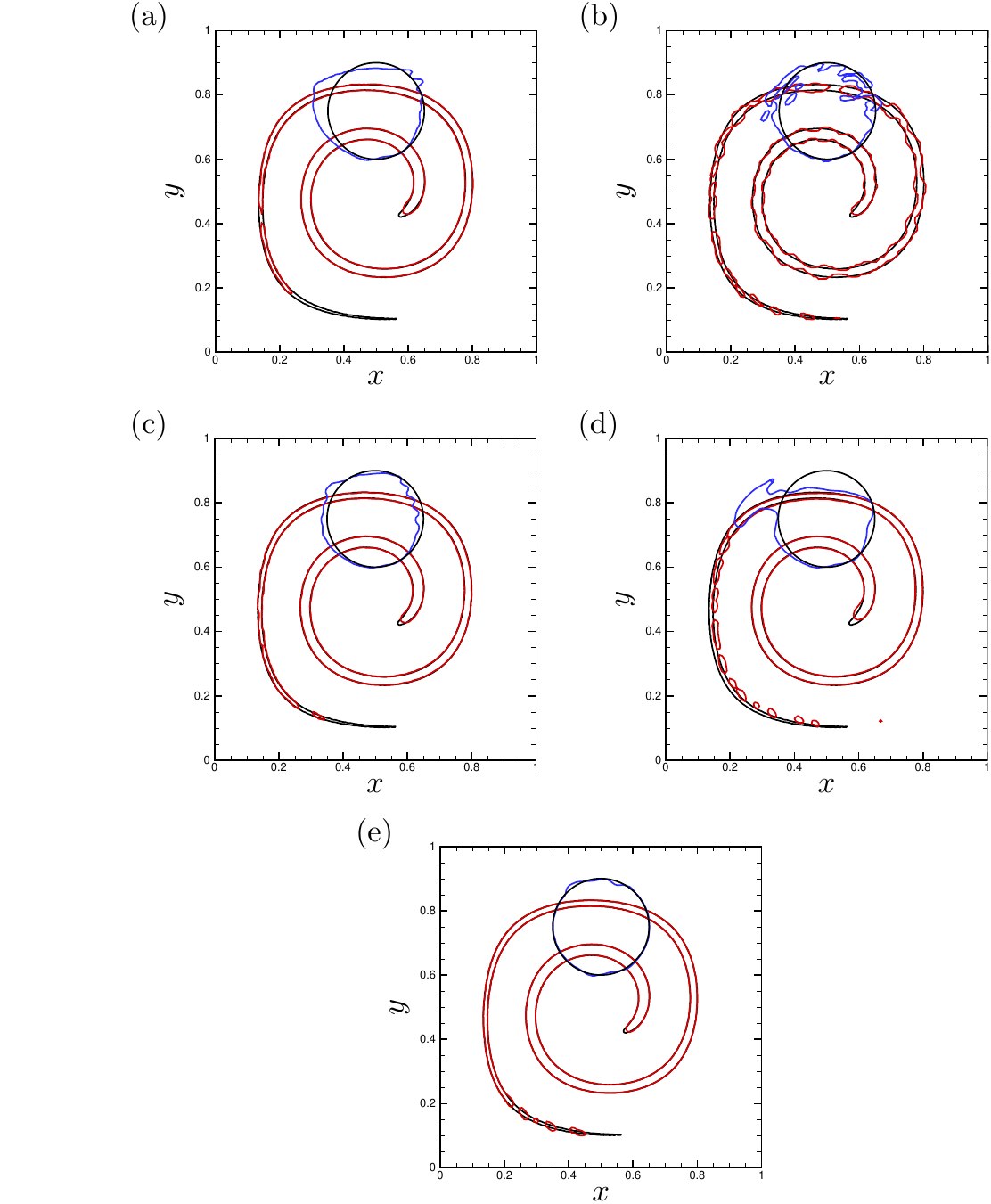} 
  \caption{Rider-Kothe reversed vortex test case: $C=0.5$ iso-line at $t=4$ (red lines) and at $t=8$ (blue lines), 
    compared with the exact results (black lines).
    Results are shown for methods SW-TVD (a), UB-TVD (b), EB-TVD (c), THINC/SW (d), PLIC (e).}
  \label{fig:RRK}
 \end{center}
\end{figure}

\begin{figure}
 \centering
 \includegraphics[]{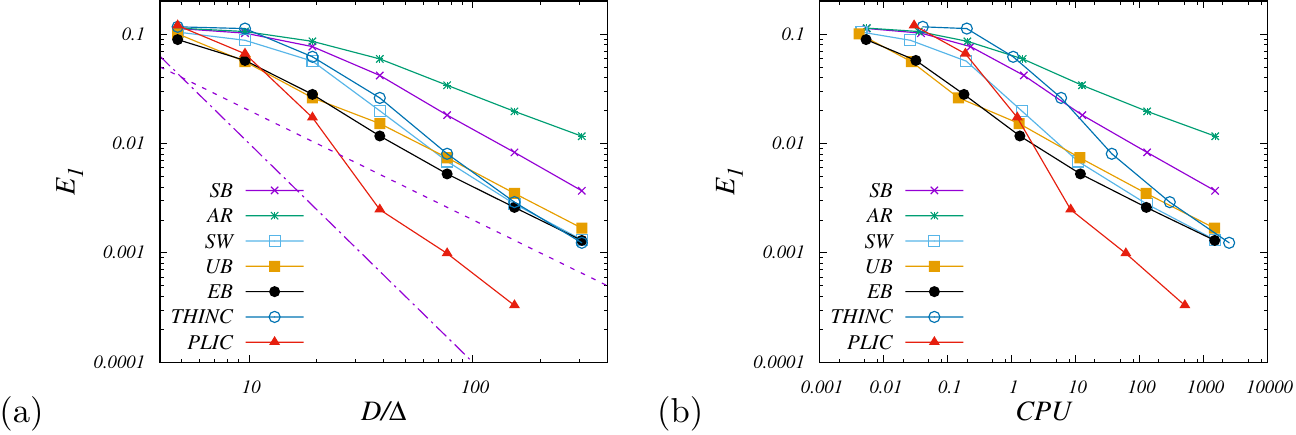} 
 \caption{Rider-Kothe reversed vortex test case: $L_1$ error norm at $t=8$ as a function of mesh resolution (a), and as a function of CPU time (in seconds). The dashed and dash-dotted lines in panel (a) denote the $(D/\Delta)^{-1}$ and $(D/\Delta)^{-2}$ trends, respectively, with $D$ the disk diameter and $\Delta$ the mesh spacing.}
 \label{fig:err_RRK}
\end{figure}

This is a two-dimensional test case in which a circle with diameter $D=0.3$ is initially centered at $(0.50, 0.75)$ in a unit square computational domain~\citep{rider_98}. This is advected by a time-varying velocity field defined by 
\begin{equation}
\left\{
\begin{array}{ccc} 
u & = & - 2 \sin^2 (\pi x) \sin (\pi y) \cos (\pi y) \cos (\pi t / T), \cr
v & = & \,\,\,\,  2 \sin^2 (\pi y) \sin (\pi x) \cos (\pi x)  \cos (\pi t / T),
\end{array}
\right.
\end{equation}
with $T=8$.
The domain is partitioned with $128^2$ uniform cells, and all boundaries are periodic. The tracer field is integrated in time and the initially circular patch is stretched and spirals around the center of the domain. Numerical results are presented at $t=4$, corresponding to maximum interface shearing, and at $t=8$, 
at which the initial conditions should be exactly reproduced by an ideal scheme.
Hence, comparison of the final and initial conditions provides a measure for the reversibility 
of numerical interface transport algorithms.
The numerically computed interface shapes are shown in Fig.~\ref{fig:RRK},
where for comparison we also show a reference solution at $t=4$ obtained with $512^2$ grid points.
Severe atomization of the spiral tail is observed at $t=4$ for the UB-TVD and THINC/SW schemes,
corresponding to an overcompressive behavior. As a result, return to the initial condition at $t=8$
is far from perfect, with severe displacement and distortion of the nominally circular patch.
The SW limiter yields less fragmentation, however at the expense of extra diffusion of the spiral tail,
whereas the EB limiter has an intermediate behavior, qualitatively comparable with the PLIC results.
PLIC however retains clear superiority over all other methods as far as return to the initial conditions
is concerned.

The error analysis for this test case is reported in Fig.~\ref{fig:err_RRK},
where results for various schemes at various mesh resolutions are presented.
Similar to the Zalesak test case, the EB limiter is found to yield the most accurate
results among TVD schemes for given mesh resolution. Comparable results are obtained 
with THINC/SW on fine meshes, whereas PLIC yields by far the most accurate results.
When the higher computational cost is considered (see panel (b)),
lower efficiency of PLIC is observed in low-fidelity computations, and higher
efficiency in highly resolved computations, the cross-over occurring at
relative errors of about $1\%$.

\subsection{Enright test case}

\begin{figure}[]
 \begin{center}
  \includegraphics[]{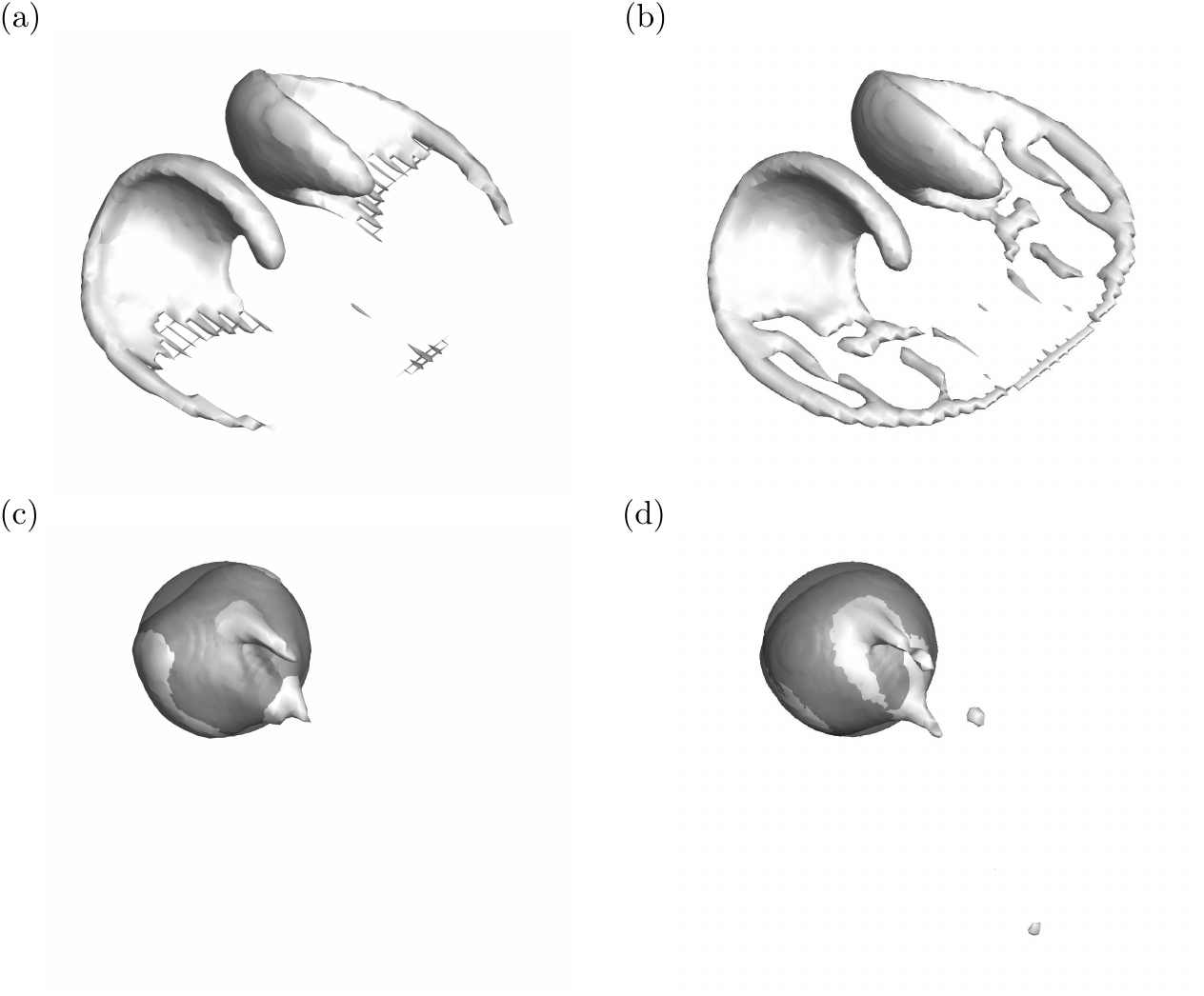} 
  \caption{Enright test case: $C=0.5$ iso-surface computed with $64^3$ grid at $t=1.5$ (top row) and $t=3$ (bottom row) for methods EB (left column), and PLIC (right column). For reference, the initial spherical patch shape is shown in panels (c), (d) in dark shades.}
  \label{fig:enright_64}
 \end{center}
\end{figure}

\begin{figure}[]
 \begin{center}
  \includegraphics[]{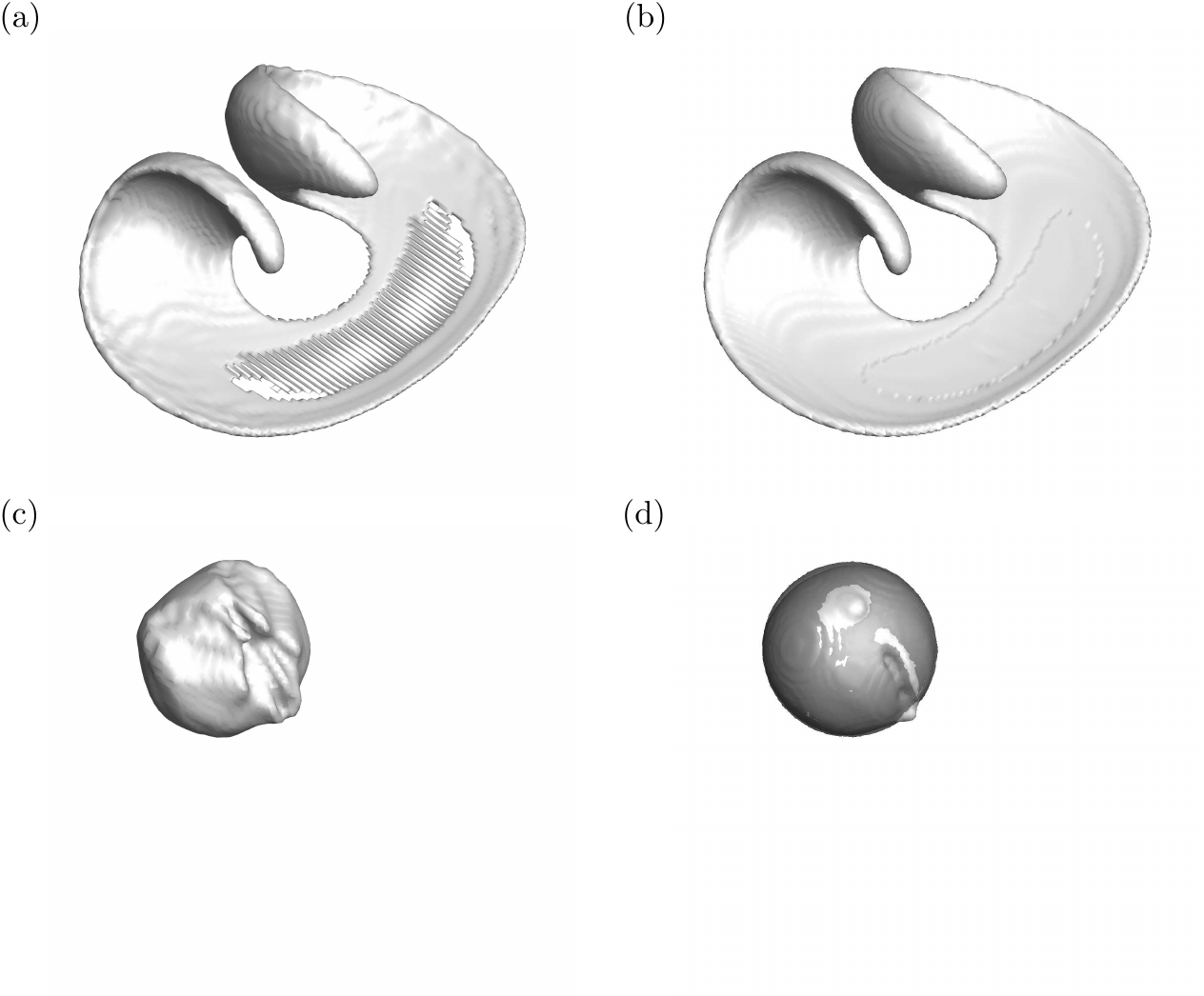} 
  \caption{Enright test case: $C=0.5$ iso-surface computed with $128^3$ grid at $t=1.5$ (top row) and $t=3$ (bottom row) for methods EB (left column), and PLIC (right column). For reference, the initial spherical patch shape is shown in panels (c), (d) in dark shades.}
  \label{fig:enright_128}
 \end{center}
\end{figure}

\begin{figure}
 \centering
 \includegraphics[]{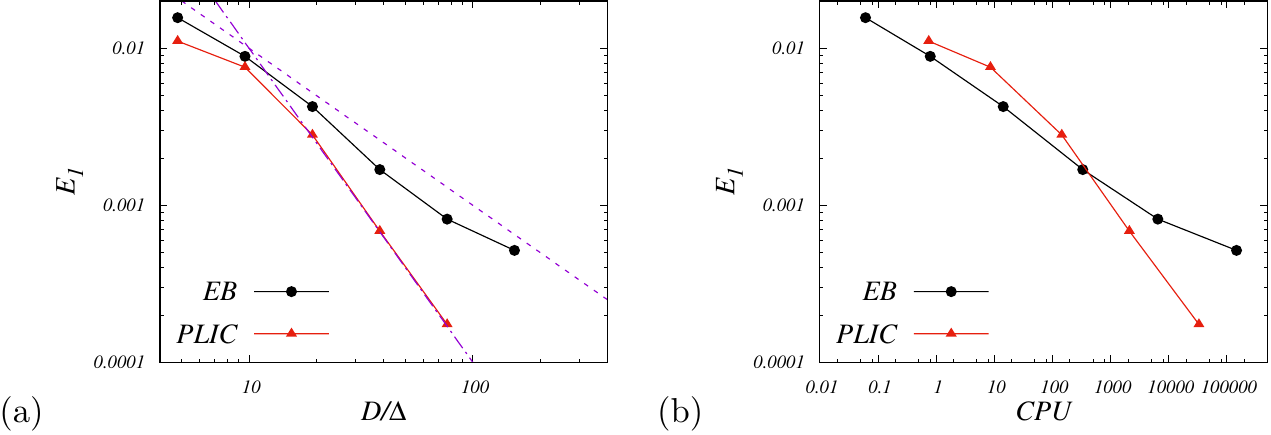} 
 \caption{Enright test case: $L_1$ error norm at $t=3$ as a function of mesh resolution (a), and as a function of CPU time (in seconds). The dashed and dash-dotted lines in panel (a) denote the $(D/\Delta)^{-1}$ and $(D/\Delta)^{-2}$ trends, respectively, with $D$ the sphere diameter and $\Delta$ the mesh spacing.}
 \label{fig:err_Enright}
\end{figure}

In this three-dimensional test case~\citep{enright_02}, a sphere with diameter $D=0.3$ and centered at $(0.35, 0.35, 0.35)$ is advected by a time-varying velocity field
\begin{equation}
\left\{
\begin{array}{ccc} 
u &=& 2 \sin^2 (\pi x) \sin (2 \pi y) \sin (2 \pi z) \cos (\pi t / T), \cr
v &=& - \sin (2 \pi x) \sin^2 (\pi y) \sin (2 \pi z) \cos (\pi t / T), \cr
w &=& - \sin (2 \pi x) \sin (2 \pi y) \sin^2 (\pi z) \cos (\pi t / T),
\end{array}
\right.
\end{equation}
with period $T=3$. This flow stretches the sphere into a thin sheet creating two bending and
spiralling tongues. Maximum deformation is reached at $t = 1.5$, at which the temporal cosine prefactor
completely quenches the sphere, making it difficult for numerical methods to resolve the interface. 
From here on the flow reverses, and the interface should return to its initial shape and position at time $t = 3$.
Numerical results at $t=1.5$ and at $t=3$ are shown in Figs.~\ref{fig:enright_64} and \ref{fig:enright_128} on $64^3$ and $128^3$ grids, respectively, limited to the EB-TVD and PLIC schemes. At the coarser $64^3$ resolution, both schemes
fail to capture the severe thinning of the interface, which is broken up into two separate chunks, although it should be noted that the interface is visually smoother in PLIC. At the final time the initial shape is qualitatively retained by both schemes, albeit with some trailing filaments, more evident in the EB-TVD scheme, and with some floatsam in PLIC. At the higher $128^3$ resolution, partial transient loss of resolution is still observed at the intermediate time from EB-TVD, which also exhibits slight wrinkling of the interface. PLIC doesn't show any tearing here, and yields better return to the initial spherical interface shape.
The quantitative error analysis reported in Fig.~\ref{fig:err_Enright} again shows significantly lower error of PLIC for given mesh resolution, and asymptotically second-order convergence. However, when the different computational effort of each simulation is accounted for (we find that the grind time of PLIC is about a factor six higher than TVD-VOF in these three-dimensional experiments), the EB-TVD scheme proves superior computational efficiency at affordable computational cost, whereas PLIC shows superiority when operated on fine meshes.

\section{Conclusions}
\label{sec:conclusions}

Numerical experiments of interface transport have been carried out by means of 
algebraic TVD schemes which entirely avoid multi-dimensional reconstruction.
Among this class of schemes we have identified a new, CFL-dependent limiter (here labelled as extra-bee, EB)
which compromises between extreme numerical anti-diffusion conveyed by downwinding, and
classical shock-capturing limiters. We have shown that the EB-TVD scheme 
supports asymptotic square wave profiles including at most two transition points,
and it yields minimum error for all test cases herein reported within the class
of TVD schemes. Nearly identical performance to the widely used algebraic THINC scheme
is found in one-dimensional cases, and similar performance in multi-dimensional
cases, at lower computational cost.
Comparison with geometric PLIC schemes in a recent efficient implementation~\citep{Basilisk}
shows, as expected, higher numerical error on a given mesh. 
Especially apparent is numerical anisotropy which yields large errors in the 
case of smooth interfaces. Tests carried out for more challenging two- and three-dimensional 
test cases featuring extreme thinning of the interface however show comparable
error on coarse meshes, whereas PLIC retains clear advantage on finer meshes.
Accounting for the large computational cost saving (which we find to be of about a 
factor two over THINC, and a factor of four to six over PLIC) 
shows that algebraic TVD schemes offer may offer comparable computational 
efficiency (i.e. lower cost for given error level, or vice-versa),
at least when computational resources are limited and/or error tolerance is not too strict.
Higher order PLIC however retains advantage in the range of highly resolved simulation.

Several concluding remarks should be made. 
First, the computational cost of material interface transport is typically only a fraction
of a full multi-phase Navier-Stokes solver, hence the advantage of improved (or similar) 
efficiency as classical PLIC solvers only partly carries over to the overall computational cost.
Second, direct numerical simulation of multi-component Navier-Stokes equations including spray formation involves
simultaneous resolution of a broad range of interfacial flow scales on the same mesh, and the main practical challenge
is the design of a mesh which has a barely sufficient number of points per bubble radius
that the smallest energetically relevant bubbles are adequately resolved.
If that is the case, the larger, more energetic bubbles are also certainly well resolved.
This amounts to say that (relatively) low-accuracy schemes as those under scrutiny here,
which bear advantage for poorly resolved flow structures can perform quite well
in practical DNS, as is actually the case for the single-component Navier-Stokes equations~\citep{orlandi_12}.
Potential improvements of the simple method herein presented might include 
improvement in accuracy through coupling with the level-set method, for instance
in the setting recently proposed by \citet{qian_18}.

\section*{Acknowledgment}

This work has been financially supported by the Flagship Project RITMARE ``The Italian Research for the Sea'', coordinated by the Italian National Research Council and funded by the Italian Ministry of Education, University and Research within the National Research Program 2014-2015. 
The authors wish to thank P.L. Roe for bringing to our attention an
insightful paper~\citep{roe_85}.

\Urlmuskip=0mu plus 1mu\relax
\bibliographystyle{model1-num-names}
\bibliography{references.bib}

\end{document}